# Absence of ferromagnetism in Mn- and Co-doped ZnO

C. N. R. Rao [*] and F. L. Deepak

Chemistry and Physics of Materials Unit and CSIR Centre of Excellence in Chemistry, Jawaharlal Nehru Centre for Advanced Scientific Research, Jakkur PO, Bangalore 560064, India.

**Abstract**

Following the theoretical predictions of ferromagnetism in Mn- and Co-doped ZnO, several workers reported ferromagnetism in thin films as well as in bulk samples of these materials. While some observe room-temperature ferromagnetism, others find magnetization at low temperatures. Some of the reports, however, cast considerable doubt on the magnetism of Mn- and Co-doped ZnO. In order to conclusively establish the properties of Mn- and Co-doped ZnO, samples with 6 % and 2 % dopant concentrations, have been prepared by the low-temperature decomposition of acetate solid solutions. The samples have been characterized by x-ray diffraction, EDAX and spectroscopic methods to ensure that the dopants are substitutional. All the Mn- and Co-doped ZnO samples (prepared at 400 ºC and 500 ºC) fail to show ferromagnetism. Instead, their magnetic properties are best described by a Curie-Weiss type behavior. It appears unlikely that these materials would be useful for spintronics, unless additional carriers are introduced by some means.

*For correspondence: Fax: 91-80-22082760, E-mail: cnrrao@jncasr.ac.in

**Introduction**

Materials for spintronics are receiving increasing attention in the last few years. A variety of materials, specially diluted magnetic semiconductors, have been investigated in this connection.[1,2] In the year 2000, Dietl et al.[3] made the theoretical prediction that Mn-doped ZnO and GaN would be ferromagnetic at room temperature and would therefore be suitable for applications in spintronics. A report of ferromagnetism in Co-doped $TiO_2$[4] gave the hope that Co- and Mn-doped oxides may indeed be useful for spintronics. Theoretical calculations of Sato and Katayama-Yoshida[5] showed that ZnO doped with several 3d transition metal ions such as V, Cr, Fe, Co and Ni may exhibit ferromagnetic ordering. A number of workers have, therefore, investigated ZnO doped with transition metal ions in the last few years, in particular, thin films of Co- and Mn-doped ZnO. Results of these studies have been reviewed by Chambers and Farrow [6] and Prellier et al.[7] The latter authors conclude that the Co-doped ZnO films generally exhibit ferromagnetism above room temperature, and that a definitive $T_C$ is not always found in Mn-doped ZnO thin films. Results from the recent literature, however, reveal many contradictions. Thus, thin films of $Zn_{1-x}Mn_xO$ (x = 0.1 and 0.3) grown on $Al_2O_3$ substrates by laser MBE are reported to show a $T_C$ in the 30-45 K range by Jung et al,[8] but Fukumura et al.[9] find a spin-glass behavior with strong antiferromagnetic exchange coupling in similar films. A first principles study of $Zn_{1-x}Mn_xO$ thin films predicts the coupling between the Mn ions to be antiferromagnetic.[10] Studies have been reported on bulk samples of Mn-doped ZnO as well. Han et al.[11] report a ferrimagnetic phase transition in the case of $Zn_{0.95}Mn_{0.05}O$ processed at 1170 K which they attribute to the $(Mn,Zn)Mn_2O_4$ spinel impurity. Such a transition was not found in samples prepared at 1370 K. Polycrystalline and single crystalline $Zn_{1-x}Mn_xO$ samples have not been found to be ferromagnetic.[12] Mn-doped ZnO nanowires prepared at high temperatures by carbon-assisted synthesis in this laboratory were only paramagnetic. The recent report of room-temperature ferromagnetism in both bulk and thin films of $Zn_{1-x}Mn_xO$ (x = 0.01, 0.02 and 0.1) by Sharma et al.[13] has aroused much interest. These workers prepared their samples at relatively low temperatures and observed weak ferromagnetism ($T_C$ > 420 °C) with an average magnetic moment of 0.16 µB per ion. The samples prepared at higher temperatures (> 700 °C) did not exhibit ferromagnetism. Sharma et al. prepared their samples by mixing ZnO and $MnO_2$ powders and calcining the mixture at 400 °C or above. The state of Mn in such preparations seems rather uncertain. Clearly, the present situation with regard to the magnetic properties of Mn-doped ZnO is far from being clear.

The experimental situation of Co-doped ZnO is similar to that of Mn-doped ZnO. Although earlier work on Co-doped ZnO films showed them to be ferromagnetic with a $T_C$ > 280 K,[14] recent results are not altogether conclusive. $Zn_{1-x}Co_xO$ films obtained by the sol-gel method were found to be ferromagnetic with a $T_C$ > 300 K by Lee et al.[15] although the presence of a secondary phase was noted in the samples x ≥ 0.25. Films of Co-doped ZnO prepared by pulsed laser deposition are reported to be ferromagnetic at room temperature,[16] but Norton et al.[17] suggest that Co nanocrystallites present in the sample could be responsible for the ferromagnetism. Room-temperature ferromagnetism and negative magnetoresistance were reported recently by Yan et al.[18] on thin films synthesized on the subnanometer scale by sputtering. Polycrystalline monophasic samples do not appear to exhibit ferromagnetism.[12a] Risbud et al.[19] show that well-characterized stoichiometric bulk samples of $Zn_{1-x}Co_xO$ are not ferromagnetic and indicate dominant nearest-neighbour antiferromagnetic interaction. These samples were prepared by heating a solid solution of zinc and cobalt oxalates at a temperature of 1173 K for 15 min. Deka et al.[20] have reported ferromagnetism upto 750 K in polycrystalline $Zn_{1-x}Co_xO$ prepared by combustion synthesis. The reaction temperature in such a combustion synthesis will be rather high. Colloidal $Co^{2+}$ doped ZnO nanocrystals prepared by the isocrystalline core-shell method are reported to be ferromagnetic.[21] Interestingly, $Zn_{1-x}Mn_xO$ (x = 0.05 - 0.1) nanocrystals prepared under solvothermal conditions by the decomposition of the cupferron precursor was only paramagnetic.[22] A first principles study has, however, shown that Co-doped ZnO prefers to be in a spin-glass state due to antiferromagnetic superexchange interactions.[23]

The above discussion demonstrates how the occurrence of ferromagnetism in bulk as well as in thin films of Mn- and Co-doped ZnO is by no means established. A careful examination of the published papers indicates that where ferromagnetism has been found, the samples were heated to relatively high temperatures which could give rise to spinel impurity phases. Risbud et al,[19] however, report the absence of ferromagnetism in samples prepared at high temperatures. Even where the temperature of synthesis is relatively low, some of the synthetic procedures are not convincing as to whether the dopant has substituted the Zn site. While it is possible that Co clusters may be present in some of the Co-ZnO samples due to the reduction of $Co^{2+}$ (that can occur in solution phase even at low temperatures), the presence of the magnetic spinel phases cannot be entirely eliminated in some of the preparations. Furthermore, the magnetization values reported by many workers is very low and can arise from the presence of magnetic impurities which cannot be detected by x-ray diffraction.

Considering the situation described above, it seemed desirable to investigate the magnetic properties of Mn- and Co-doped ZnO prepared at low temperatures, ensuring that the dopant ions are present in substitutional sites. This has been accomplished by preparing the doped ZnO samples by the thermal decomposition (330 °C) of the solid solutions of Zn(acetate)$_2$ with Mn(acetate)$_2$ and Co(acetate)$_2$ and subjecting to them all possible means of chracterization. We have carried out careful studies on samples with 6 % Mn and 6 % Co, as well as 2 % Mn and 2 % Co. The latter was necessary since the proponents of ferromagnetism state that the dopant concentration has to be low ($\leq 4$ %) for observing ferromagnetism.[13, 24] Samples prepared at 400 °C and 500 °C have been studied to avoid any doubtful conclusions that may arise because of the temperature employed for the sample preparation. Interestingly, we find that both the 2 % and 6 % Mn- and Co-doped ZnO samples fail to exhibit ferromagnetism.

**Experimental**

Preparation of Mn- doped ZnO:

Zinc acetate, $(CH_3COO)_2$ Zn.2H$_2$O, and manganese acetate, $(CH_3COO)_2$ Mn.4H$_2$O, supplied by Aldrich were taken in the required molar ratios (6 mole % or 2 mole % Mn) and dissolved in 15 ml double distilled water. It was made sure that the solution did not have precipitates before drying at 100 °C overnight (~ 10 h). The powder obtained after drying was heated in air at 400 °C for 5 h. (This sample is referred to as I). Another sample was prepared by heating the product of decomposition of the acetate solid solutions at 500 °C for 5 h. (This sample is referred to as II). Heating and cooling rates in both the cases were 1 °C/min. The product obtained was light brown in colour in both I and II.

Preparation of Co- doped ZnO:

Zinc acetate and cobalt (II) acetate, $(CH_3COO)_2$ Co.$4H_2O$, supplied by Aldrich was taken in the required molar ratios (6 mole % and 2 mole %) and dissolved in 15 ml double distilled water. It was made sure that the solution did not have any precipitate before drying at 100 °C overnight (~ 10 h). The powder obtained after drying was heated in air at 400 °C for 5 h (sample I). Another sample was prepared by heating the product of decomposition of the acetate solid solutions at 500 °C for 5 h. (sample II) Heating and cooling rates in both the cases were 1 °C/min. The product obtained was dark green in colour in both these cases.

Thermogravimetric analysis (TGA) of the Zn-Mn and Zn-Co acetate solid solutions was carried out on a Mettler-Toledo-TG-850 instrument. X-ray diffraction (XRD) patterns were recorded using a Seifert (XRD, XDL, TT, and Cu target) instrument. The chemical composition was determined with an Oxford EDX analyzer attached with a Leica S-440i SEM instrument. Transmission electron microscopy was carried out with a JEOL JEM 3010 instrument operating at an accelerating voltage of 300 kV. X-ray photoelectron spectra of the samples were recorded with an ESCALAB MKIV spectrometer employing AlKα radiation (1486.6 eV). Electronic absorption spectra were recorded in the 2000-200 nm range using a Perkin-Elmer Lambda 900 UV/VIS/NIR spectrophotometer. Photoluminescence (PL) measurements were carried out with a Perkin-Elmer LS 50B luminescence spectrophotometer with an excitation wavelength of 325 nm. Electron paramagnetic resonance (EPR) spectra were recorded with a ER 200 D X-Band Bruker instrument. Magnetic properties of the various powder samples were measured using a SQUID magnetometer (Quantum Design MPMS) which has a base temperature of 2 K and a maximum magnetic field of 5 T.

**Results and Discussion**

Thermogravimetric analysis curves of the Zn-Mn and Zn-Co acetate solid solutions showed that they decompose sharply around 330 °C, giving doped ZnO as the product (Fig. 1). We have characterized the Mn- and Co-doped ZnO samples prepared by the decomposition of the acetate solid solutions at 400 °C (I) and 500 °C (II) by employing various techniques. X-ray diffraction patterns of the products of decomposition of the acetate solid solutions (Fig. 2) showed the hexagonal structure, the Mn-doped samples exhibiting a slightly larger c-parameter (6 % Mn-doped, a = 3.250 Å, c = 5.224 Å; 6 % Co-doped, a = 3.249 Å, c = 5.207 Å and 2 % Mn-doped, a = 3.250 Å, c = 5.220 Å; 2 % Co-doped, a = 3.249 Å, c = 5.206 Å) in comparison to that of the undoped sample (a = 3.249 Å, c = 5.206 Å). The error of fit was generally around 0.5. The increase in the c-parameter results from the substitution of $Mn^{2+}$ ions because of the larger radius of $Mn^{2+}$ (0.66 Å) compared to $Zn^{2+}$ (0.60 Å).[8] On the other hand, the radius of $Co^{2+}$ (0.58 Å) is close to that of $Zn^{2+}$ and as a result the cell parameters do not vary significantly.[19] The absence of impurity peaks arising from secondary phases or precipitates implies that the percentage doping employed is within the solubility limits of Mn or Co in ZnO. That the solubility limit of $Mn^{2+}$ and $Co^{2+}$ is far greater than the percentages employed by us is well-documented.[11,15] Energy dispersive X-ray (EDAX) analysis (see typical data in Fig. 3) confirmed the concentrations of Mn and Co to be close to those in the starting nominal compositions. Transmission electron microscope examination showed the doped samples to consist of particles of 30-50 nm diameter. X-ray photoelectron spectroscopy revealed that the Mn and Co ions were in the +2 oxidation state. Thus, the Mn (2p) and Co (2p) signals were found at 641.45 eV and 778.93 eV respectively.[15]

The Mn-doped ZnO samples gave a broad absorption band in the 400-450 nm region due to $^6A_1$ (S) → $^4T_1$ (G) transition (Fig. 4). The Mn-doped sample also gave the characteristic EPR spectrum of $Mn^{2+}$ (Fig. 5) with a g value of 2.003, consistent with that reported in single crystals of Mn-doped bulk ZnO.[12b] The Co-doped ZnO samples gave three bands in the 550-700 nm region due to the $^4A_2$ (F) → $^2E$ (G) (659 nm), $^4A_2$ (F) → $^4T_1$ (P) (615 nm) and $^4A_2$ (F) → $^2A_1$ (G) (568 nm) transitions characteristic of the tetrahedral $Co^{2+}$ ions.[25] (Fig. 4) The optical energy gaps of the doped ZnO samples were considerably smaller, showing thereby that the band gap of ZnO can be tuned by such doping.[8] Both the doped samples gave UV as well as blue-green emissions, somewhat weaker than in undoped ZnO.[26,27] With the various characterization data mentioned above, we conclude that both $Mn^{2+}$ and $Co^{2+}$ ions are present substitutionally in the $Zn^{2+}$ sites of ZnO, in the samples prepared by us.

Detailed magnetic measurements were carried out on the 6 % and 2 % Mn (Co)-doped ZnO samples. We show the temperature variation of the inverse susceptibility $\chi^{-1}_M$ of the Mn-doped samples in Fig. 6 and the Co-doped samples in Fig. 7. The zero-field cooled (ZFC) and field cooled (FC, 1000 Oe) data are comparable indicating that the material does not possess the characteristics of a spin glass. Extrapolation of the inverse susceptibility data in the high temperature region gave negative Curie temperatures of – 5 and – 15 K for Mn-doped samples I and II respectively; the values were – 65 and – 15 K for Co-doped I and II samples. The data can be understood in terms of the model of Spalek et al.[28] employing a modified Curie-Weiss law, or the model of Lawes et al.[29] wherein magnetization is treated as a sum of a Curie-Weiss term and a Curie term with a large Weiss temperature. Furthermore, we do not observe magnetic hysteresis or any other evidence for ferromagnetic ordering down to 2 K. The results show that the 6 % Mn- and 6 % Co-doped ZnO prepared by us exhibit only antiferromagnetic superexchange interactions but no ferromagnetism.

After completing our work on 6 % Mn- and 6 % Co-doped samples, some workers claimed that it was necessary to have lower percentage of dopants ( < 4 %) to observe ferromagnetism.[13,24] We, have therefore, carried out investigations of 2 % Mn- and Co-doped ZnO samples. (see insets of Figures 3 and 4). In Fig. 8 (a) and (b), we show the temperature variation of the inverse susceptibility $\chi^{-1}_M$ of the 2 % Mn- and Co-doped ZnO samples (I). The ZFC and FC data show little difference. The data in the high-temperature region give paramagnetic Curie temperatures of 5 K and 50 K for the Mn- and Co- doped samples, but the M vs H curves show no hysteresis (Fig. 9).

**Conclusions**

The present investigations on Mn- and Co-doped ZnO establish them not to be ferromagnetic and throw considerable doubt about the ferromagnetic nature of these materials reported in the literature. It seems unlikely that these materials would be candidates for spintronics. This conclusion finds support from the recent work of Spaldin[30] who finds that robust ferromagnetism cannot occur in Mn- and Co-doped ZnO. If at all, it may occur if additional charge carriers are present. In order to obtain robust ferromagnetism, it may be worthwhile to investigate the effect of codoping of Mn- or Co-doped ZnO samples with other cations to induce additional charge carriers.

**Acknowledgements**: The authors thank DRDO (India) and Department of Science and Technology for support of this research. The authors thank Drs. Ram Seshadri and K. Ramesha for magnetic measurements and Dr. A. Govindaraj for assistance in synthesis.

**References**

1  S. A. Chambers and Y. K. Yoo, *MRS Bulletin*, 2003, **28**(10), 706 - 707.

2  M. Ziese and M. F. Thornton, eds., *Spin Electronics* (Springer, Berlin, 2001).

3  T. Dietl, H. Ohno, F. Matsukura, J. Gilbert and D. Ferrand, *Science*, 2000, **287**(5455), 1019 - 1022.

4  Y. Matsumoto, M. Murakami, T. Shono, T. Hasegawa, T. Fukumura, M. Kawasaki, P. Ahmet, T. Chikyow, S.-Y. Koshihara and H. Koinuma, *Science*, 2001, **291**(5505) 854 - 856.

5  K. Sato and H. Katayama-Yoshida, *Japan. J. Appl. Phys.*, 2000, **39**, L555 – L558.

6  S. A. Chambers and R. F. C. Farrow, *MRS Bulletin*, 2003, **28**(10), 729 - 733.

7  W. Prellier, A. Fouchet and B. Mercey, *J. Phys. Condens. Matter*, 2003, **15**, R1583 - R1601.

8  S. W. Jung, S. -J. An, G-C Yi, C. U. Jung, S. Lee and S. Cho, *Appl. Phys. Lett.*, 2002, **80**(24), 4561 - 4563.

9  T. Fukumura, Z. Jin, M. Kawasaki, T. Shono, T. Hasegawa, S. Koshihara and H. Koinuma, *Appl. Phys. Lett.*, 2001, **78**(7) 958 - 960.

10  Q. Wang P. Jena, *Appl. Phys. Lett.*, 2004, **84**(21) 4170 - 4172.

11  S-J. Han, T. - H. Jang, Y. B. Kim, B. –G. Park, J. –H. Park and Y. H. Jeong, *Appl. Phys. Lett*, 2003, **83**(5) 920 - 922.

12  (a) S. Kolesnik, B. Dabrowski and J. Mais, *Phys. Stat. Sol.*, 2004, **1**(4) 900 - 903; (b) V. Yu. Ivanov, M. Godlewski, S. Yatsunenko, A. Khachapuridze, Z. Golacki, M. Sawicki, A. Omelchuk, M. Bulany and A. Gorban, *Phys. Stat. Sol.* 2004, **1**(2) 250 - 253.

13  P. Sharma, A. Gupta, K. V. Rao, F. J. Owens, R. Sharma, R. Ahuja, J. M. Osorio Guillen, B. Johansson, G. A. Gehring, *Nature Mater.*, 2003, **2**, 673 - 677.

14  K. Ueda, H. Tabata and T. Kawai, *Appl. Phys. Lett.*, 2001, **79**(7), 988 - 990.

15  H-J. Lee, S-Y. Jeong, C. R. Cho and C. H. Park, *Appl. Phys. Lett.*, 2002, **81**(21), 4020 - 4022.

16  (a) K. Rode, A. Anane, R. Mattana, J. –P. Contour, O. Durand and R. LeBourgeois, *Appl. Phys. Lett.*, 2003, **93**(10), 7676 - 7678; (b) S. Ramachandran, A. Tiwari, J. Narayan, *Appl. Phys. Lett.*, 2004, **84**(25), 5255 - 5257.

17  D. P. Norton, M. E. Overberg, S. J. Pearton, K. Pruessner, J. D. Budai, L. A. Boatner, M. F. Chisholm, J. S. Lee, Z. G. Khim, Y. D. Park, R. G. Wilson, *Appl. Phys. Lett.*, 2003, **83**(26), 5488 - 5490.

18  S-s. Yan, C. Ren, X. Wang, Y. Xin, Z. X. Zhou, L. M. Mei, M. J. Ren, Y. X. Chen, Y. H. Liu and H. Garmestani, *Appl. Phys. Lett.*, 2003, **84**(13), 2376 - 2378.

19  A. S. Risbud, N. A. Spaldin, Z. Q. Chen, S. Stemmer and R. Seshadri, *Phys. Rev. B.*, 2003, **68**, 205202-1 - 205202-7.

20  S. Deka, R. Pasricha, P. A. Joy, *Chem. Mater.*, 2004, **16**, 1168 - 1169.

21  D. A. Schwartz, N. S. Norberg, Q. P. Nguyen, J. M. Parker, D. R. Gamelin, *J. Am. Chem. Soc.*, 2003, **125**, 13205 - 13218.

22  M. Ghosh, R. Seshadri, C. N. R. Rao, *J. Nanosci. Nanotech*, 2004, **4**(1/2), 136 - 140.

23  E –C. Lee and K. J. Chang, *Phys. Rev. B.*, 2004, **69**, 085205-1 – 085205-5.

24  H. J. Blythe, R. M. Ibrahim, G. A. Gehring, unpublished results.

25  F. A. Cotton, D. M. L. Goodgame, M. Goodgame, *J. Am. Chem. Soc.*, 1961, **83**, 4690 - 4699.

26  M. Liu, A. H. Kitai, P. Mascher, *Jl. of Luminescence*, 1992, **54**, 35 - 42.

27  X. T. Zhang, Y.C .Liu, J. Y. Zhang, Y. M. Lu, D.Z. Shen, X. W. Fan, X. G. Kong, *J. Crystal Growth*, 2003, **254**, 80 - 85.

28  J. Spalek, A. Lewicki, Z. Tarnawski, J. K. Furdyna, R. Galazka and Z. Obuszko, *Phys. Rev. B*., 1986, **33**(5), 3407 - 3418.

29  G. Lawes, A. P. Ramirez, A. S. Risbud and Ram Seshadri, arXiv:cond-mat/0403196 v1, 2004.

30  N. A. Spaldin, *Phys. Rev. B*., 2004, **69**, 125201-1 – 125201-7.

**Figure Captions**

**Fig. 1** TGA plots of (a) Mn-Zn and (b) Co-Zn acetate solid solutions.

**Fig. 2** XRD patterns of (a) 6 % Mn-doped and (b) 6 % Co-doped ZnO.

**Fig. 3** EDAX spectra of (a) 6 % Mn-doped and (b) 6 % Co-doped ZnO. Insets show the corresponding EDAX spectra for the 2 % Mn- and Co-doped ZnO.

**Fig. 4** Absorption spectra of (a) ZnO, (b) 6 % Mn-doped and (c) 6 % Co-doped ZnO. Inset shows the spectrum of 2 % Co-doped ZnO.

**Fig. 5** EPR spectrum of the 6 % Mn-doped ZnO.

**Fig. 6** Temperature variation of inverse magnetic susceptibility of 6 % Mn-doped ZnO heated to 400 ºC (I). Inset shows the data for sample II heated to 500 ºC.

**Fig. 7** Temperature variation of inverse magnetic susceptibility of 6 % Co-doped ZnO heated to 400 ºC (I). Inset shows the data for sample II heated to 500 ºC.

**Fig. 8** Temperature variation of inverse magnetic susceptibility of (a) 2 % Mn-doped and (b) 2 % Co-doped ZnO.

**Fig. 9** M vs H plots of (a) 2 % Mn-doped and (b) 2 % Co-doped ZnO obtained at an external field of T = 5 K.

# Figure: 1

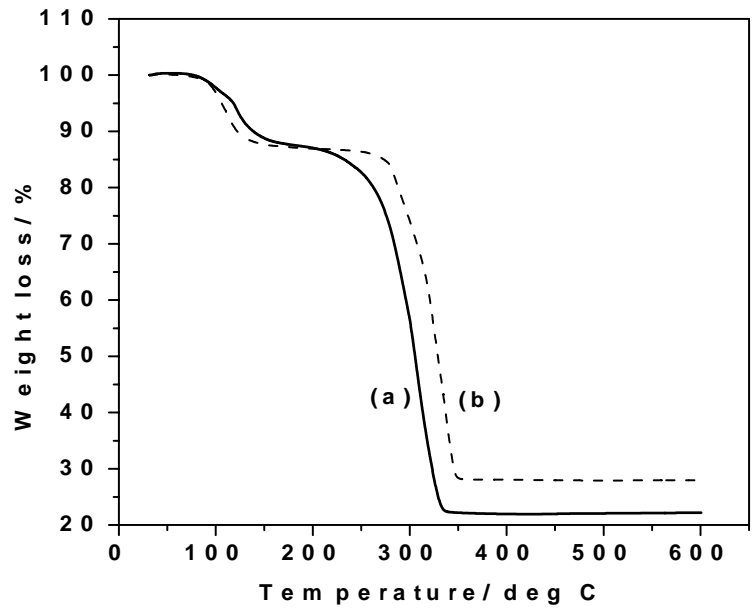

**Figure: 2**

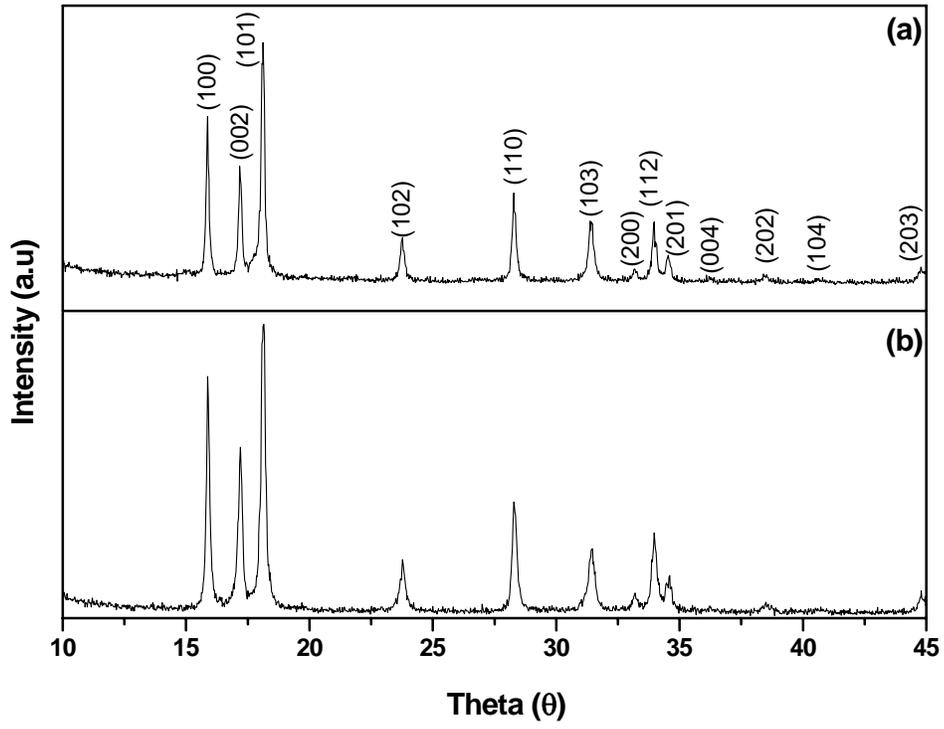

**Figure: 3**

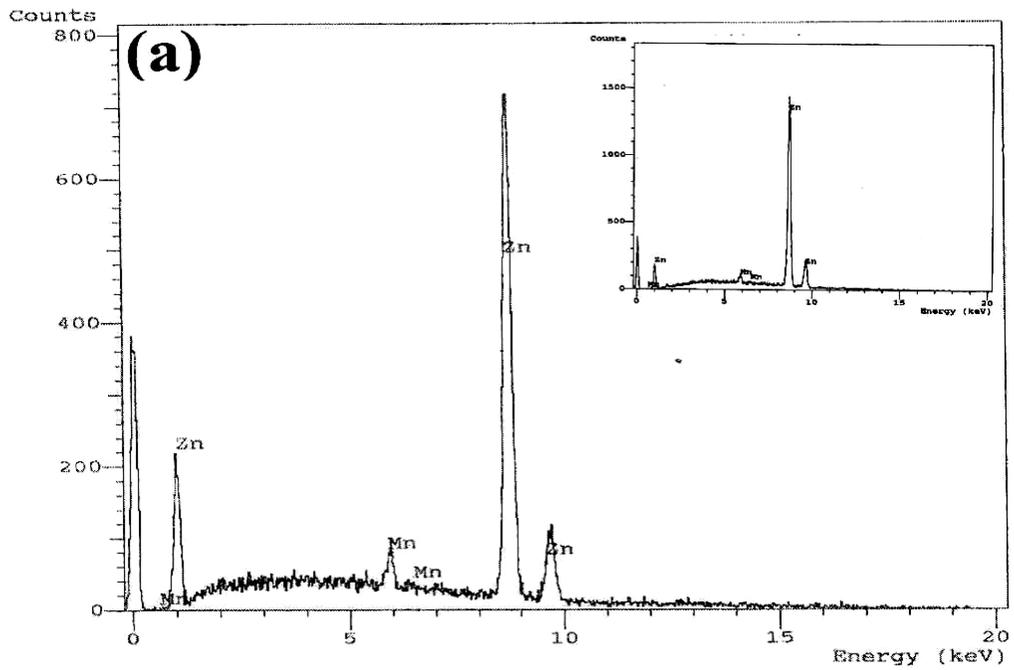

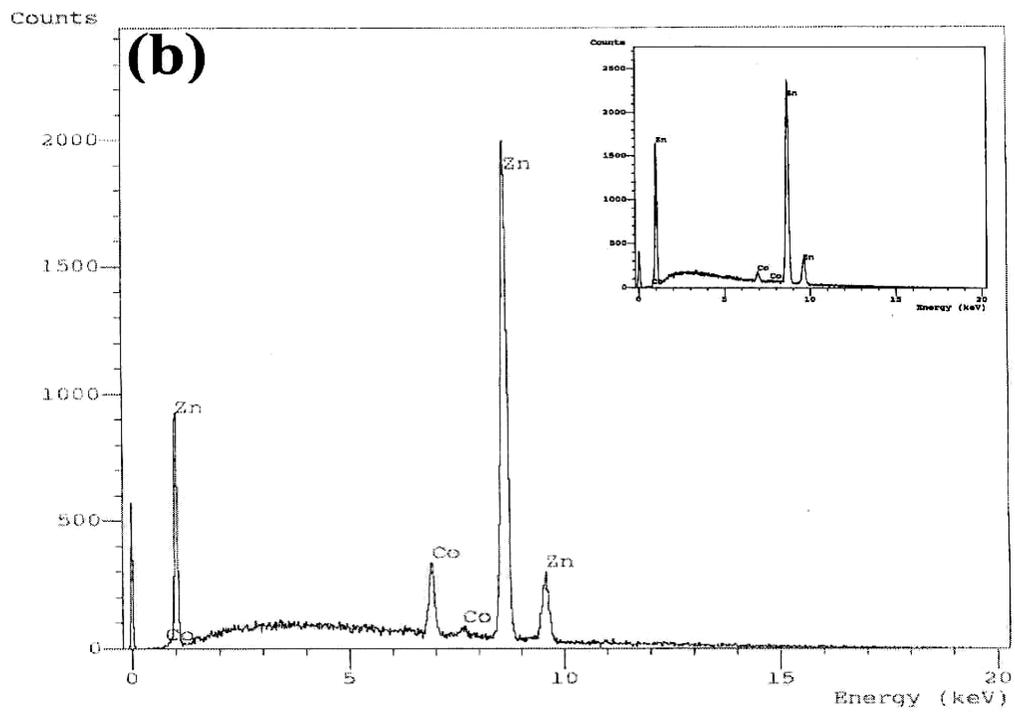

**Figure: 4**

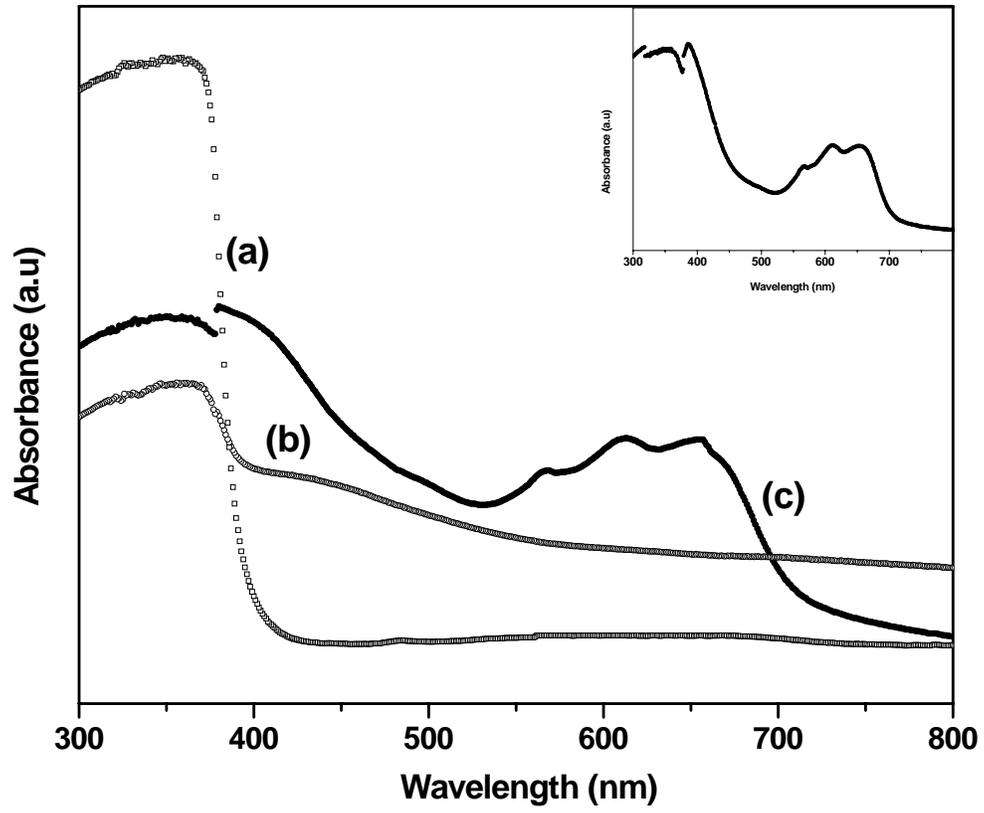

**Figure: 5**

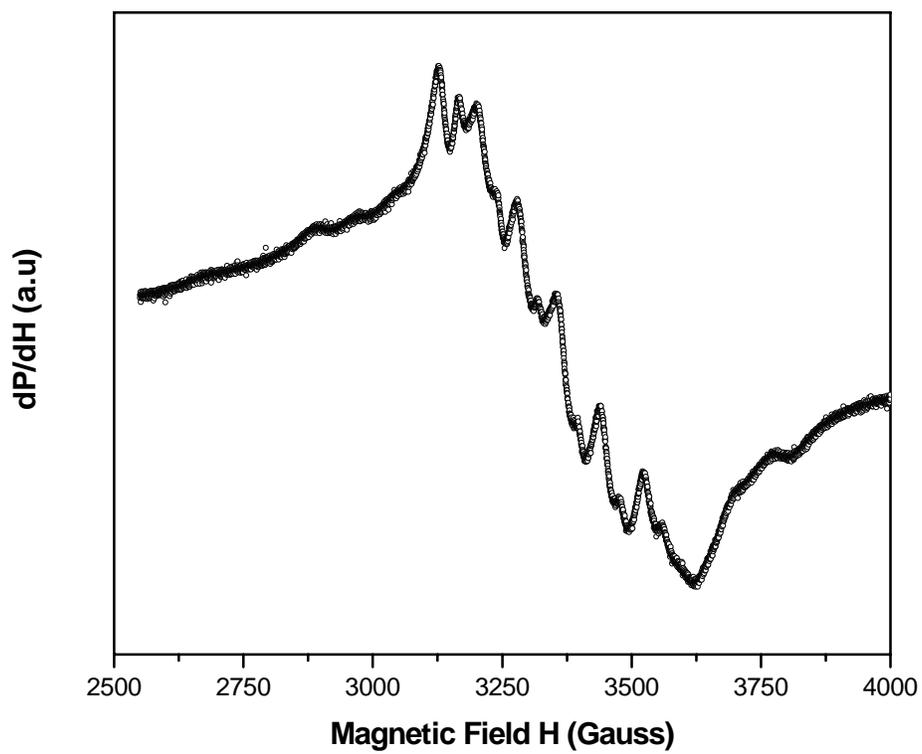

# Figure: 6

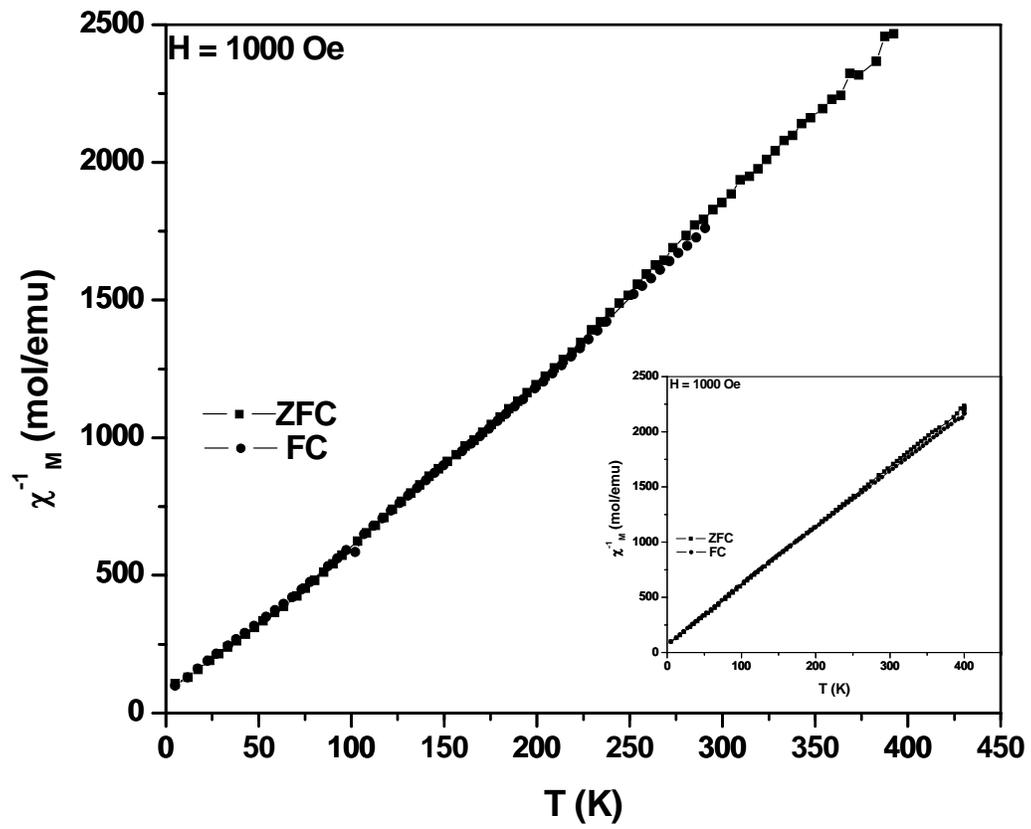

# Figure: 7

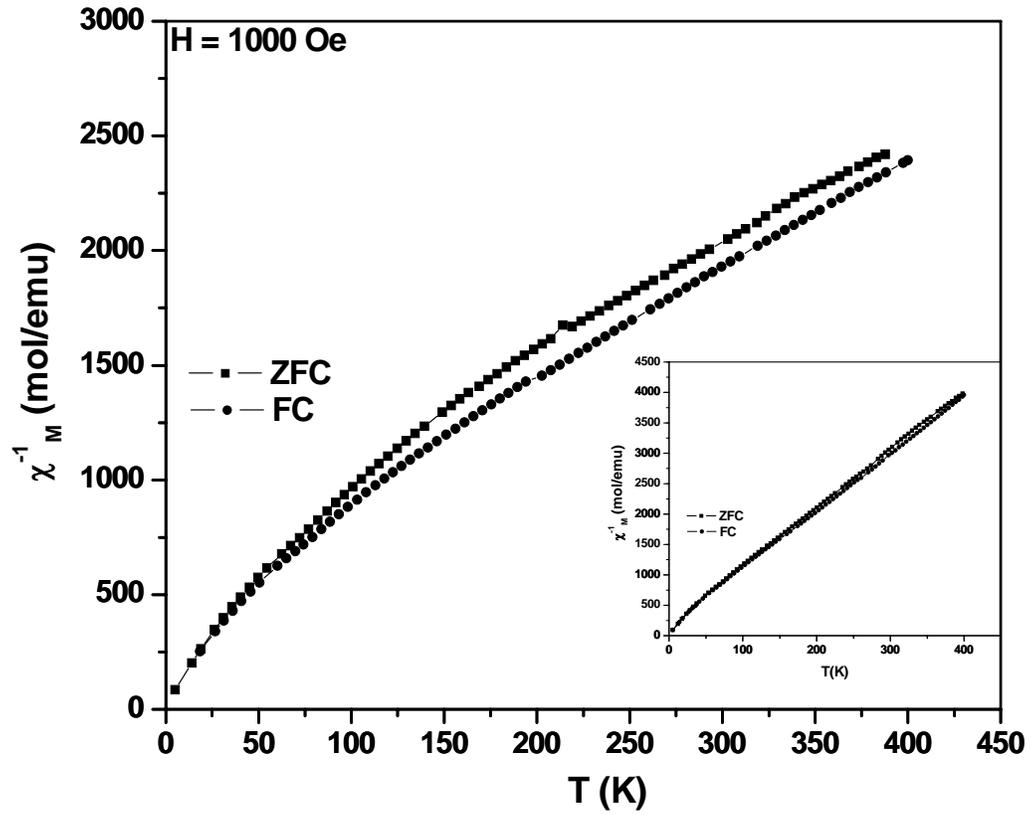

# Figure: 8

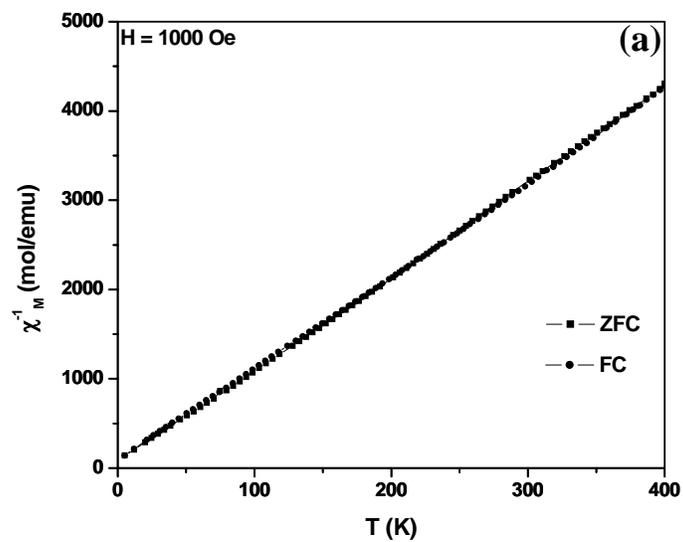

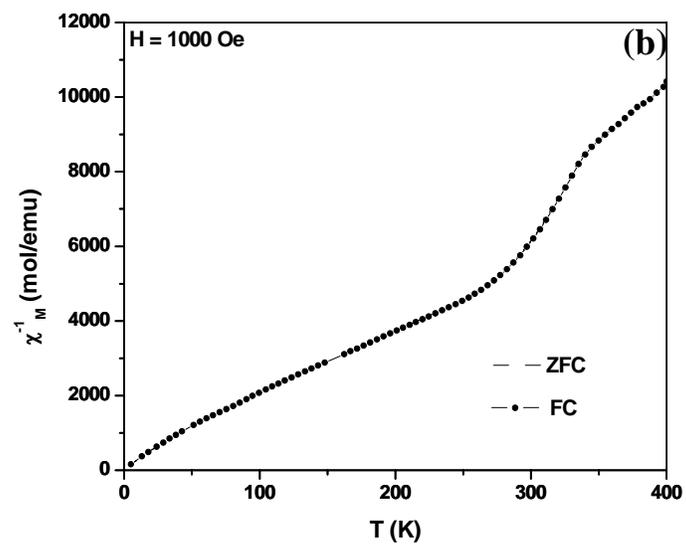

**Figure: 9**

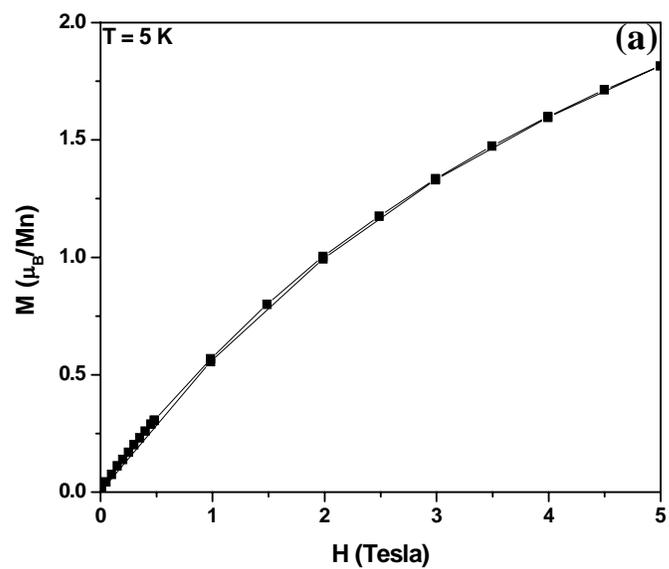

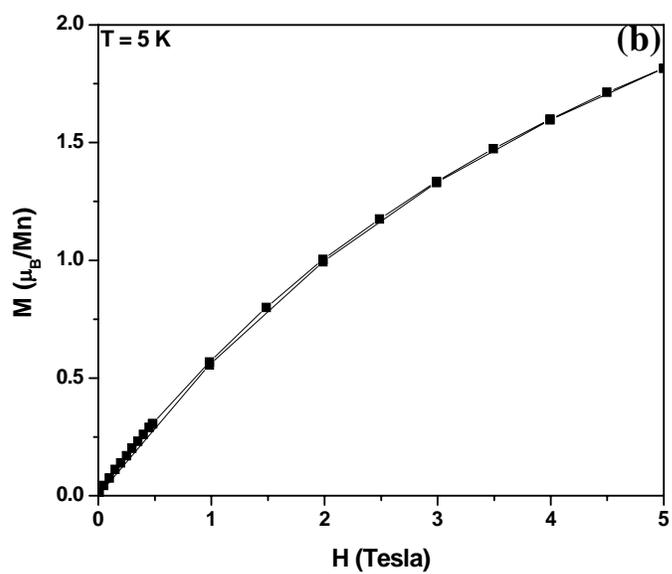